# Improving BIM Authoring Process Reproducibility with Enhanced BIM Logging

Suhyung Jang[1] Ghang Lee, Ph.D.[2]

[1]Building Informatics Group, Dept. of Architecture and Architectural Engineering, Yonsei Univ. ORCID: https://orcid.org/ 0000-0001-8289-0657. Email: rgb000@yonsei.ac.kr.
[2]Professor, Dept. of Architecture and Architectural Engineering, Yonsei Univ. (corresponding author). ORCID: https://orcid.org/ 0000-0002-3522-2733. Email: glee@yonsei.ac.kr

**ABSTRACT**

This paper presents an enhanced building information modeling (BIM) logger that captures building element geometry and attributes to accurately represent the BIM authoring process. The authors developed the logger and reproducing algorithm using the Revit C# API based on the analysis of information required to define building elements and associated attributes. The enhanced BIM log was evaluated through a case study of Villa Savoye designed by Le Corbusier, and the results show that it can accurately represent the BIM authoring process to a substantial level of reproducibility. The study provides a tool for capturing and reproducing the BIM authoring process. Future research can focus on improving the accuracy of the logging and reproducing algorithm and exploring further applications to automate the BIM authoring process using enhanced BIM logs.

**INTRODUCTION**

The widespread adoption of BIM technology has transformed the architecture, engineering, construction, and operation (AECO) industry. Although BIM data is widely used to support decision-making in AECO project management, the decision-making process during BIM authoring is not captured in the final BIM models. BIM log mining studies have attempted to understand the BIM authoring process by utilizing the BIM log, which contains sequential records of events recorded during BIM software usage. Previous studies have focused on various aspects of the BIM authoring process, such as design authoring patterns (Yarmohammadi et al. 2017), collaboration patterns (Zhang and Ashuri 2018), and the role of modelers (Forcael et al. 2020), but the absence of element-centric information items limits the reproducibility of the BIM log to better represent the BIM authoring process (Jang et al. 2021).

To address this issue, previous studies have proposed custom BIM loggers, but they do not capture enough information to reproduce the BIM authoring process accurately. Therefore, this paper presents an enhanced BIM logger that captures sufficient information to accurately reproduce the BIM authoring process. The paper describes the methodology employed to develop the enhanced logger, which includes analyzing the minimum information requirements to define building elements in Autodesk Revit, developing a custom BIM logger that captures the necessary information, and developing a reproducing algorithm to evaluate the effectiveness of the enhanced logger in accurately representing the BIM authoring process. The reproducibility of the enhanced BIM log was evaluated through a case study.

The paper is structured as follows. Section 2 reviews previous custom BIM loggers proposed in the literature, and section 3 describes the research methodology employed in this study. Section 4 reviews the minimum information requirements to define BIM elements in Revit, and section 5 outlines the development of the enhanced BIM logger and reproducing algorithm. Section 6 evaluates the reproducibility of the enhanced BIM logger through a case study, and section 7 concludes the paper.

**LITERATURE REVIEW**

BIM log mining is a data analysis approach that utilizes process mining techniques to explore BIM event logs collected during a BIM software operation. Process mining includes various techniques for automated process discovery, social network analysis, process optimization, case prediction, and history-based recommendations (Aalst et al. 2011). However, event log imperfections can lead to unreliable results, and researchers have introduced an incremental approach to evaluating event log fitness and a methodology to guide process mining execution (Bose et al. 2013; Suriadi et al. 2017).

While improving event log quality has received significant attention, several studies have focused on improving the quality of BIM logs. Previous research has developed logging systems to extract building element information using predefined event handlers on the Revit API (Yarmohammadi and Castro-Lacouture 2018), IFC loggers to capture BIM model updates (Kouhestani and Nik-Bakht 2020; Pan and Zhang 2021), and command-object graphs to notate the modeling process sequence (Gao et al. 2021). However, the custom logs developed in these studies also lack the sufficient information to reproducing the BIM authoring process due to missing element geometry (Jang et al. 2021).

The BIM authoring process is defined as a process in which 3D software is used to develop a BIM model based on criteria that are important to the translation of the building's design (Messner et al. 2019). Accordingly, this process includes the addition, deletion, and modification of the geometry of building elements and their associated properties (Kouhestani and Nik-Bakht 2020; Lin and Zhou 2020). Throughout the design phases, building elements are developed to meet the level of development (LOD) of the BIM model. Therefore, understanding how building elements are developed throughout the BIM authoring process is crucial for gaining insights into the process. In this regard, the objective of this study is to develop a custom logger that captures the necessary information to reproduce the BIM authoring process, which will be outlined in the following section.

**METHODOLOGY**

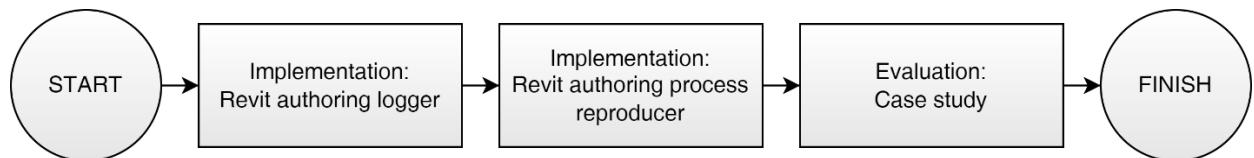

**Figure 1. Research flow**

This study developed a three-phase methodology to enhance the reproducibility of BIM logs in a case study, as depicted in Figure 1. The methodology involved developing a customized Revit authoring logger, creating a Revit authoring process reproducer, and assessing the reproducibility of the enhanced BIM log in a case study. The study utilized Autodesk Revit 2023 and Revit C# API for plugin development.

To capture sufficient information in the BIM log to reproduce the BIM authoring process, the authors analyzed the minimum inputs required to represent building elements in Revit. As building elements are represented using classes in an object-oriented programming model, the information required for each class corresponding to a specific category of elements was analyzed. The log captures string-described geometric bases and BuiltInParameters to represent the building elements in comma-separated value (CSV) format. The reproducing algorithm developed in this study iterates through the events recorded in the BIM log, identifies the command type (i.e., "ADDED," "MODIFIED," or "DELETED"), and executes them with reference to the "Comments" property (i.e., copied element ID) of the building elements.

The evaluation phase of this study involved modeling Villa Savoye designed by Le Corbusier for the case study, during which the Revit authoring logger recorded the authoring process. The events recorded in the enhanced BIM log were iterated using the Revit authoring process reproducer, and the reproducibility of the developed model was evaluated based on the average distance and volume differences between elements, as well as visual analysis of plans, elevations, sections, and 3D views.

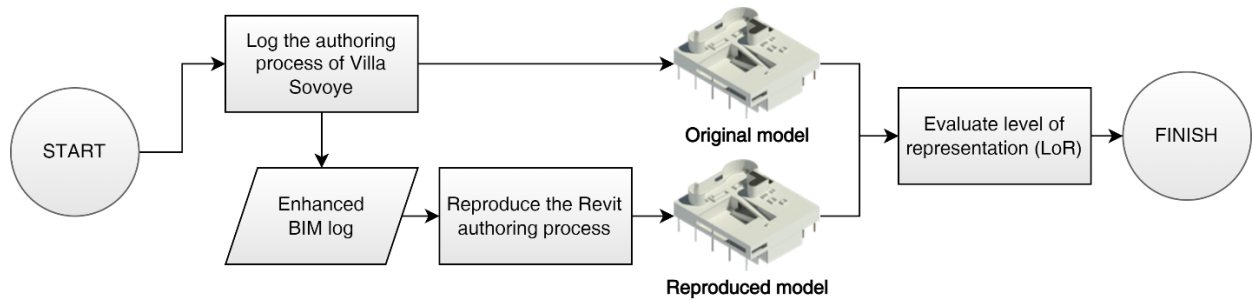

**Figure 2. Evaluation of Revit authoring logger and Revit authoring log reproducer**

**ENHANCED BIM LOGS**

This section provides a detailed overview of how building elements are defined in Revit and how the enhanced BIM log captures the necessary information to accurately represent the BIM authoring process. Building elements in Revit are represented using classes in an object-oriented programming (OOP) model that has a hierarchical structure that reflects the physical structure of the building. Each class corresponds to a specific type of building element, such as walls, floors, roofs, doors, and windows. Figure 3 illustrates the Revit API methods for defining each building element category.

Wall elements in Revit are classified into two categories—rectangular profile walls and nonrectangular profile walls—with different creation methods for each type. Floor elements are classified into two types, flat floors and sloped floors, and are created using different parameters, depending on the type. Windows, doors, and columns are classified as FamilyInstances, and their representation is based on the placement of predefined instances on the selected base geometry. FamilyInstances have different categories based on the FamilySymbol used, such as WindowType,

DoorType, and ColumnType. Windows and doors require a LocationPoint and HostElement, while columns can be placed without a HostElement, depending on whether the representation of the slope is required.

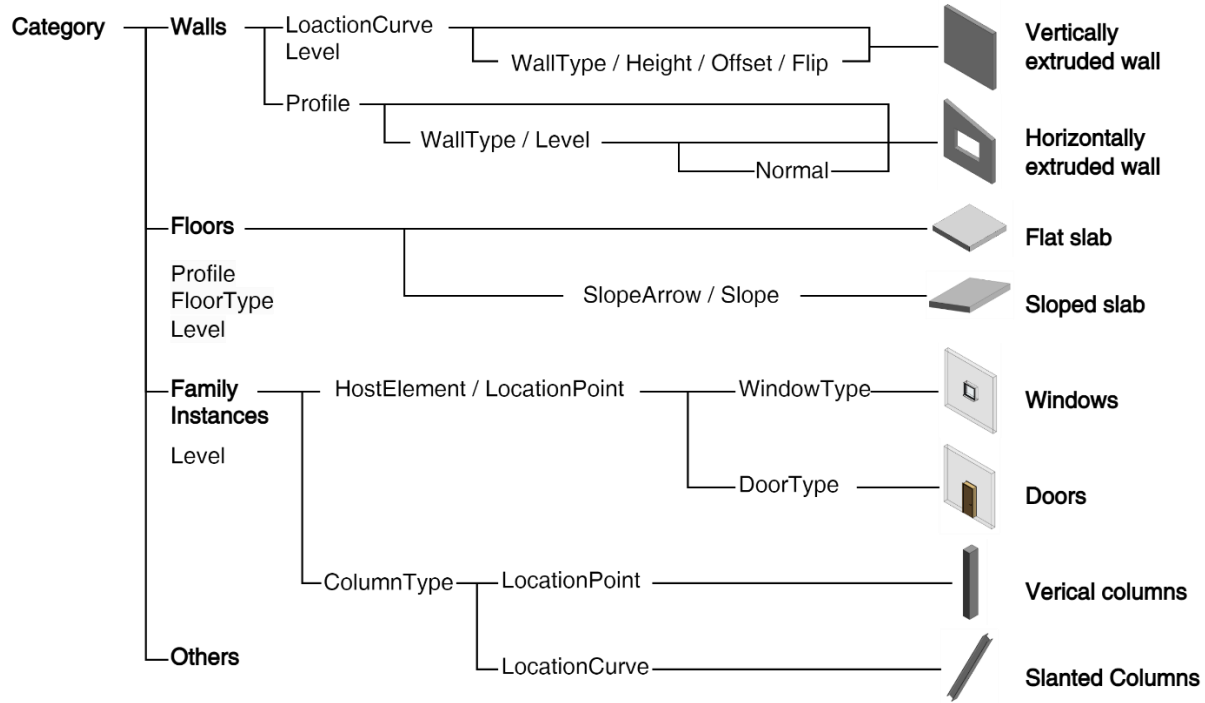

**Figure 3. Revit API methods for adding building elements by category**

In addition to recording other required information items as string formats, the enhanced BIM log captures geometric bases, such as LocationPoint, LocationCurve, and Profile, and their respective subclasses, such as Line, Arc, CylindricalHelix, Ellipse, HermiteSpline, and NurbsSpline. The information required for each geometric base and its string format representation is presented in Table 1. The enhanced BIM logger records the geometric bases of the building elements represented during the BIM authoring process in the described format.

Furthermore, there are multiple BuiltInParameters in Revit that can be modified to better represent the building element. BuiltInParameters are predefined in order of walls, floors, windows, doors, and columns. The initial values and modifications of the values of BuiltInParameters are also recorded in the enhanced log, providing additional information on how the building elements were defined during the BIM authoring process.

**Table 1. Definition of geometric classes**

| Classes | Subclasses | Input Requirements | String Formats |
|---|---|---|---|
| Location Point | XYZ | (X coordinate, Y coordinate, Z coordinate) | (double, double, double) |
| Location Curve | Line | (endPoint1, endPoint2) | [Line, XYZ, XYZ] |
| | Arc | (plane, radius, startAngle, endAngle) | [Arc, XYZ, double, double, double] |

|  | Cylindrical Helix | (basePoint, radius, xVector, zVector, pitch, startAngle, endAngle) | [CylindricalHelix, XYZ, double, XYZ, XYZ, double, double, double] |
|  | Ellipse | (center, xRadius, yRadius, xAxis, yAxis, startParameter, endParameter) | [Ellipse, XYZ, double, double, XYZ, XYZ, double, double] |
|  | Nurbs Spline | (degree, knots, controlPoints, weights) | [NurbsSpline, int, IList<double>, IList<XYZ>, IList<double>] |
|  | Hermite Spline | (controlPoints, periodic, tangents) | [HermiteSpline, IList<XYZ>, bool, HermiteSplineTangents] |
| CurveLoop | CurveLoop | (LocationCurve$_1$, …, LocationCurve$_n$) | {CurveLoop, LocationCurve$_1$, …, LocationCurve$_n$} |
| Profile | Profile | (CurveLoop$_1$, … ,CurveLoop$_n$) | Profile, CurveLoop$_1$, … ,CurveLoop$_n$ |

## REPRODUCING ALGORITHM

The authors implemented a reproducing algorithm to iterate through the events in the enhanced BIM log and repeat them, as illustrated in Figure 5. The algorithm begins by identifying the command type of events. If the command type is "ADDED", the algorithm adds an element of the recorded category and applies the recorded BuiltInParameter, while adding the ElementID of the event to the Comments parameter of the newly created BIM element. If the command type is "MODIFIED", the algorithm queries the element with the ElementID recorded in the Comments value and applies the corresponding modification. If the command type is "DELETED", the algorithm queries the elements with the ElementID recorded in the Comments value and deletes the element.

## EVALUATION OF REPRODUCIBILITY

To assess the enhanced log and its ability to be reproduced, the authors of this study carried out a case BIM authoring process of Villa Savoye designed by Le Corbusier, as depicted in (a) of Figure 6. The BIM model comprised 97 walls, 8 slabs, 8 windows, 19 doors, and 27 columns, while 2,836 events were recorded for the authoring process. All elements included in the original modeling process existed in the reproduced model with matching element IDs. The authors calculated the average differences between the location points of the elements, which were 3.6440E-07, 2.1470E-07, 1.4760E-07, 1.6139E-07, and 1.4380E-07 for walls, floors, windows, doors, and columns, respectively. The average volume differences by percentage (absolute of changed volumes of reproduced building elements divided by the original building element's volume) were 0.1876%, 0.0198%, 0.0000%, 0.0000%, and 0.0433% in the order of wall, floor, window, door, and column. It is assumed that the minor discrepancies resulted from the conversion between metric and imperial units.

However, there were also misrepresentations identified in the reproduced BIM model. As (b). of Figure 6 illustrates, The pair of horizontally extruded walls with different profiles. The difference resulted from the immature implementation of the logging and reproducing algorithm, and improvements are required. (c) Figure 6 depicts the differences between the BIM models by illustrating the drawings from each model in different colors. Although the study did not focus on the alignment of curtain wall mullions, there was a discrepancy in their alignment. There was also a discrepancy between the wall elements in a few cases, as depicted in the plan. Despite these discrepancies, the reproduced BIM model represents the original BIM model to a considerable

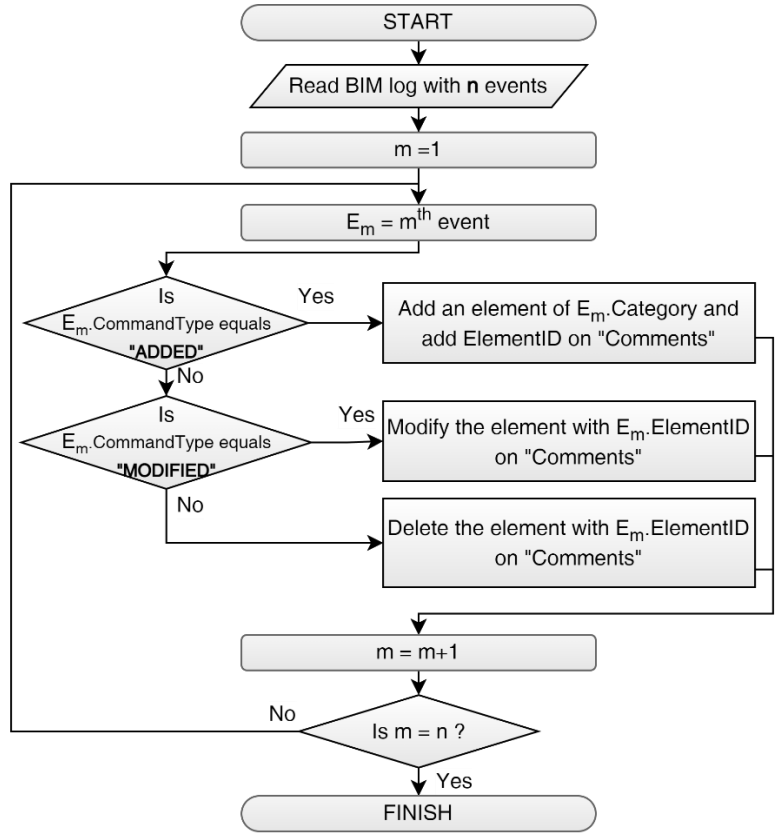

extent.

**Figure 5. Reproducing algorithm**

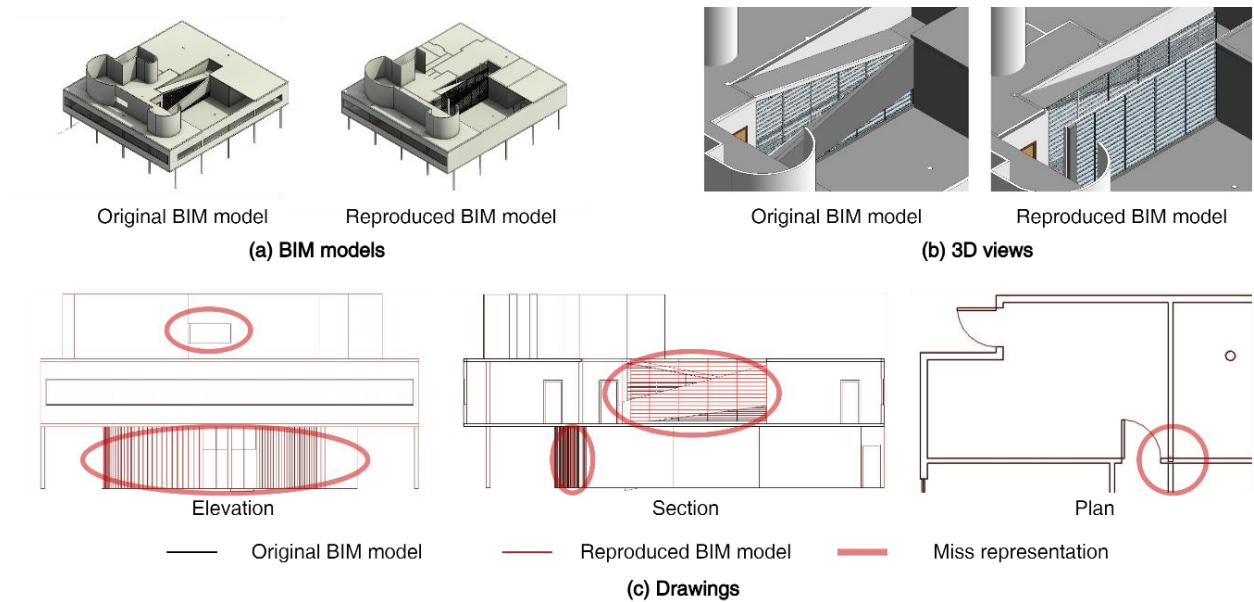

**Figure 6. Visual comparison between original BIM model and reproduced BIM model**

## CONCLUSION

In summary, this study developed an enhanced BIM logger that captures the necessary information to reproduce the BIM authoring process. By analyzing the information requirements of five representative building elements in Revit, the authors developed a custom logger that records geometric bases and BuiltInParameters. The study also developed a reproducing algorithm to repeat the BIM authoring process. The effectiveness of the enhanced log was evaluated in a case study of Villa Savoye designed by Le Corbusier, which showed that the enhanced BIM logger provides a valuable tool for capturing and reproducing the BIM authoring process to a considerable degree. While minor discrepancies and misrepresentations were observed, the results of the case study demonstrated the potential of the enhanced BIM logger.

Future research can focus on improving the accuracy of the logging and reproducing algorithm, exploring further applications of the enhanced BIM log, and investigating the potential to automate the BIM authoring process using enhanced BIM logs. Overall, the enhanced BIM logger presented in this study can contribute to improving the efficiency and accuracy of BIM authoring processes, enabling better collaboration among stakeholders and enhancing the quality of BIM models.

## ACKNOWLEDGMENTS

This work is supported by the Korea Agency for Infrastructure Technology Advancement (KAIA) grant funded by the Ministry of Land, Infrastructure and Transport (Grant 23AATD-C163269-03).

## REFERENCES

Bose, R. J. C., R. S. Mans, and W. M. van der Aalst. 2013. "Wanna improve process mining results?" *2013 IEEE Symposium on Computational Intelligence and Data Mining (CIDM)*, 127–134. IEEE.

Forcael, E., A. Martinez-Rocamora, J. Sepulveda-Morales, R. Garcia-Alvarado, A. Nope-Bernal, and F. Leighton. 2020. "Behavior and Performance of BIM Users in a Collaborative Work Environment." *Appl. Sci.-Basel*, 10 (6): 2199. Basel: Mdpi. https://doi.org/10.3390/app10062199.

Gao, W., C. Wu, W. Huang, B. Lin, and X. Su. 2021. "A data structure for studying 3D modeling design behavior based on event logs." *Automation in Construction*.

Jang, S., S. Shin, and G. Lee. 2021. "Logging Modeling Events to Enhance the Reproducibility of a Modeling Process." *ISARC. Proceedings of the International Symposium on Automation and Robotics in Construction*, 256–263. IAARC Publications.

Kouhestani, S., and M. Nik-Bakht. 2020. "IFC-based process mining for design authoring." *Automation in Construction*.

Lin, J.-R., and Y.-C. Zhou. 2020. "Semantic classification and hash code accelerated detection of design changes in BIM models." *Automation in Construction*.

Messner, J., C. Anumba, C. Dubler, S. Goodman, C. Kasprzak, R. Kreider, R. Leicht, C. Saluja, and N. Zikic. 2019. *BIM Project Execution Planning Guide (v. 2.2)*. Computer Integrated Construction Research Program, Pennsylvania State University.

Pan, Y., and L. Zhang. 2021. "A BIM-data mining integrated digital twin framework for advanced project management." *Automation in Construction*.

Pezeshki, Z., A. Soleimani, and A. Darabi. 2019. "Application of BEM and using BIM database for BEM: A review." *Journal of Building Engineering*, 23: 1–17. https://doi.org/10.1016/j.jobe.2019.01.021.

Suriadi, S., R. Andrews, A. H. ter Hofstede, and M. T. Wynn. 2017. "Event log imperfection patterns for process mining: Towards a systematic approach to cleaning event logs." *Information Systems*, 64: 132–150. Elsevier.

Van der Aalst, W., A. Adriansyah, A. K. A. de Medeiros, F. Arcieri, T. Baier, T. Blickle, J. C. Bose, P. van den Brand, R. Brandtjen, and J. Buijs. 2011. "Process mining manifesto." *International Conference on Business Process Management*, 169–194. Springer.

Yarmohammadi, S., and D. Castro-Lacouture. 2018. "Automated performance measurement for 3D building modeling decisions." *Automation in Construction*.

Yarmohammadi, S., R. Pourabolghasem, and D. Castro-Lacouture. 2017. "Mining implicit 3D modeling patterns from unstructured temporal BIM log text data." *Autom. Constr.*, 81: 17–24. Amsterdam: Elsevier Science Bv. https://doi.org/10.1016/j.autcon.2017.04.012.

Zhang, L., and B. Ashuri. 2018. "BIM log mining: Discovering social networks." *Autom. Constr.*, 91: 31–43. Amsterdam: Elsevier Science Bv. https://doi.org/10.1016/j.autcon.2018.03.009.
− 8 −